# Hydrophobic, hydrophilic and charged amino acids' networks within Protein

Md. Aftabuddin and S. Kundu*
Department of Biophysics, Molecular Biology & Genetics, University of Calcutta
92 APC Road, Kolkata 700009, West Bengal, India

*Address correspondence to:
Sudip Kundu
Department of Biophysics, Molecular Biology & Genetics,
University of Calcutta,
92 APC Road, Kolkata 700009,
West Bengal, India
skbmbg@caluniv.ac.in

Running title: Protein structure network.

**Abstract:**
The native three dimensional structure of a single protein is determined by the physico chemical nature of its constituent amino acids. The twenty different types of amino acids, depending on their physico chemical properties, can be grouped into three major classes - hydrophobic, hydrophilic and charged. We have studied the anatomy of the weighted and unweighted networks of hydrophobic, hydrophilic and charged residues separately for a large number of proteins. Our results show that the average degree of the hydrophobic networks has significantly larger value than that of hydrophilic and charged networks. The average degree of the hydrophilic networks is slightly higher than that of charged networks. The average strength of the nodes of hydrophobic networks is nearly equal to that of the charged network; whereas that of hydrophilic networks has smaller value than that of hydrophobic and charged networks. The average strength for each of the three types of networks varies with its degree. The average strength of a node in charged networks increases more sharply than that of the hydrophobic and hydrophilic networks. Each of the three types of networks exhibits the 'small-world' property. Our results further indicate that the all amino acids' networks and hydrophobic networks are of assortative type. While maximum of the hydrophilic and charged networks are of assortative type, few others have the characteristics of disassortative mixing of the nodes. We have further observed that all amino acids' networks and hydrophobic networks bear the signature of hierarchy; whereas the hydrophilic and charged networks do not have any hierarchical signature.

## 1. Introduction:

Network analysis is being recognized as a powerful tool to understand the complex systems. It helps us to understand the interaction among individual components and hence to characterize the whole system. Several researchers have worked to shed light on the topology, growth and dynamics of different kinds of networks including the world wide web (WWW), food webs, gene co-expression networks, metabolic networks and protein-protein interaction networks, etc (1-9).

Efforts have also been made to transform a protein structure into a network where amino acids are nodes and their interactions are edges (10-17). However, these protein structure networks have been constructed with varying definition of nodes and edges. This network approach has been used in a number of studies, such as protein structural flexibility, prediction of key residues in protein folding, identification of functional residues, residue contribution to the protein-protein binding free energy in given complexes (10-14). Several groups have also studied the protein network to understand its topology, small world properties and behaviors of long range and short range interactions of the amino acid nodes, etc (15-17).

In almost all of the previous studies on protein structure networks, protein has been considered as an unweighted network of amino acids. Very recently, we have considered the protein as a weighted network (18). This investigation has focused on degree and strength distribution, signature of hierarchy and assortative type mixing behavior of the amino acid nodes.

A protein molecule is a polymer of different amino acids joined by peptide bonds. These twenty different amino acids have different side chains and hence have different physcio chemical properties. When a protein folds in its native conformation, its native 3D structure is determined by the physico chemical nature of its constituent amino acids. Depending on the physicochemical properties, the different amino acids fall into three major classes - hydrophobic, hydrophilic and charged residues. In this context, it would be interesting to study the network structures of hydrophobic, hydrophilic and charged residues separately. We have also recently studied the hydrophobic and hydrophilic networks (19). The analysis has mainly focused on the degree, the degree distribution and small world properties. We have found that the average degree of hydrophobic node is larger than that of hydrophilic node. We have also observed the existence of small world properties in both the cases. While the hydrophobic and hydrophilic networks we have studied previously (19) are unweighted networks; the present study considers both the weighted and unweighted networks of hydrophobic, hydrophilic and charged residues' networks. We have analyzed these networks to focus on their topology including degree, strength, strength-degree relationships, clustering coefficients, shortest path length, existence of small world property and hierarchical signature, if any, and mixing behavior of the nodes. In summary, in the present work, we have studied the anatomy of hydrophobic, hydrophilic and charged residues' networks and have also performed a comparative study among them as well as with all amino acids' networks.

## 2. Methods:
### 2.1. Construction of hydrophobic (BN), hydrophilic (IN), charged (CN) and all amino acids' (AN) networks:

Primary structure of a protein is a linear arrangement of different types of amino acids in 1D space; where any amino acid is connected with its nearest neighbors through peptide bonds. But when a protein folds in its native conformation, distant amino acids in 1D chain may also come close to each other in 3D space and hence different non-covalent interactions are possible among them depending on their orientations in 3D space. Moreover, each of the twenty amino acids has different side chains and different physico chemical properties. These different twenty amino acid residues have been grouped into three major classes - hydrophobic (F, M, W, I, V, L, P, A), hydrophilic (N, C, Q, G, S, T, Y) and charged (R, D, E, H, K). Here we are interested in studying the hydrophobic, hydrophilic and charged networks within proteins.

Any network has two basic components - nodes and edges. Only the hydrophobic residues are considered as nodes of a hydrophobic network; whereas hydrophilic and charged residues are considered as the nodes of hydrophilic and charged networks respectively. If any two atoms from two different amino acids (nodes) are within a cut-off distance (5A°) the amino acids are considered to be connected or linked. The cut-off distance is within the higher cut-off distance of London-van der Waals forces (20). Further in our calculations we have not considered the interaction of any of the backbone atoms, we have only included the interactions of the side chain atoms.

Thus, in a hydrophobic network, hydrophobic residues are nodes and the possible links among them are edges. Same logic is followed to construct the other networks.

Since we have also compared the network parameters of hydrophobic, hydrophilic and charged networks with those of all amino acids' network within protein, we have also constructed the networks taking into account all amino acids without any classifications. Thus we have obtained the unweighted networks of AN, BN, IN and CN types.

Next, we shall discuss the basis of transforming the protein structure into a weighted network. When we consider a proteins' 3D structure, several atoms of any amino acid in a protein may be within the cut-off distance of several atoms of another amino acid. This results in a possible multiple links between any two amino acids. These multiple links are the basis of the weight of the connectivity, which may vary for different combinations of amino acids as well as for different orientations of them in 3D conformational space. The intensity $w_{ij}$ of the interaction between two amino acids '$i$' and '$j$' is defined as the number of possible links between the $i$ -th and the $j$ -th amino acids. Considering the intensity of interaction between any two amino acids we have constructed the weighted hydrophobic, hydrophilic, charged and all amino acids' networks.

We have collected a total of 161 protein structures from protein crystal structure data bank (21) with following criteria -
    1) Maximum Percentage identity: 25
    2) Resolution: 0.0-2.0
    3) Maximum R-value: 0.2
    4) Sequence length: 500 - 10,000
    5) Non X-ray entries: excluded
    6) CA only entries: excluded
    7) CULLPDB by chain

In some of the crystal structures, the atomic coordinates of some of the residues are missing. We have not considered those structures since they may give erroneous values of different network parameters (degree, clustering coefficient, etc). A final set of eighty-five crystal structures was taken for the calculation and analysis of network properties. We have generated the hydrophobic, hydrophilic, charged and all amino acids' networks of each of the eighty-five proteins using the three dimensional atomic coordinates of the protein structures. While all amino acid networks for each of the proteins form a single cluster; the hydrophobic, hydrophilic and charged networks, in general, have more than one subnetwork. The number of nodes of these subnetworks vary in a wider range. The subnetworks having at least thirty nodes have been collected and analyzed.

## 2.2. Network parameters:

Each of the networks has been represented as an adjacency matrix (A). Any element of adjacency matrix (A), $a_{ij}$ is given as

$a_{ij}$ = 1, if $i \neq j$ and $i$ and $j$ nodes are connected by an edge
0, if $i \neq j$ and $i$ and $j$ nodes are not connected
0, if $i = j$.

The degree of any node '$i$' is represented by $k_i = \Sigma_j a_{ij}$

The number of possible interactions between any two amino acids may vary depending on their 3D orientations and the number of atoms in their side chains. If $w_{ij}$ is the number of possible interactions between any $i$-th and $j$-th amino acids, then the strength $(s_i)$ of a node $i$ is given by $s_i = \Sigma_j a_{ij} w_{ij}$

This parameter represents the number of connectivities of any two amino acids and is thus a characteristic of weighted network. It should be clearly mentioned that the weighted network analysis depends on (1) the number of possible interactions between amino acid residues and also on (2) the energy of interactions between them. Since the total energy of interactions again depends on the total number interactions between the residues, we, for the sake of simplicity of analysis, have considered only the number of interactions between the residues.

We have determined the characteristic path length (L) and the clustering coefficient (C) of each of the network. The characteristic path length L of a network is the path length between two nodes averaged over all pairs of nodes. The clustering coefficient $C_i$ is a measure of local cohesiveness. Traditionally the clustering coefficient $C_i$ of a node '$i$' is the ratio between the total number ($e_i$) of the edges actually connecting its nearest neighbors and the total number of all possible edges between all these nearest neighbors [$k_i (k_i - 1) / 2$; if the $i$-th vertex has $k_i$ neighbors] and is given by $C_i = 2e_i / k_i (k_i - 1)$ where $e_i$ is the total number of edges actually connecting the $i$-th node's nearest neighbors. Then the clustering coefficient of a network is the average of its all individual $C_i$'s. For a random network having N number of nodes with average degree <k>, the characteristic path length ($L_r$) and the clustering coefficient ($C_r$) have been calculated using the expressions $L_r \approx lnN / ln <k>$ and $Cr \approx <k> /N$ given in (3). To examine if there is any 'small world' property in a network, we have followed Watts & Strogatz's method (3). According to them, a network has the small world property if $C >> C_r$ and $L \geq L_r$. Combining the topological information with the weight distribution of the network, Barrat *et al.* (22) have introduced an analogous parameter to C and that is known as weighted clustering coefficient, $C^w_i$. The weighted clustering coefficient, $C^w_i$ takes into account the importance of the clustered structure on the basis of amount of interaction intensity (number of possible interactions between amino acids) actually found on the local triplets and is given by $C^w_i = [1 / s_i (k_i - 1)] \Sigma_{j,h} (w_{ij} + w_{ih}) a_{ij} a_{ih} a_{jh} / 2$.

To study the tendency for nodes in networks to be connected to other nodes that are like (or unlike) them, we have calculated the Pearson correlation coefficient of the degrees at either ends of an edge. For our undirected unweighted protein network its value has been calculated using the expression suggested by Newman (23) and is given as

$r = (M^{-1}\Sigma_i j_i k_i - [M^{-1}\Sigma_i 0.5(j_i + k_i)]^2) \div (M^{-1}\Sigma_i 0.5(j_i^2 + k_i^2) - [M^{-1}\Sigma_i 0.5(j_i + k_i)]^2)$

Here $j_i$ and $k_i$ are the degrees of the vertices at the ends of the $i$-th edge, with $i = 1, ..M$. The networks having positive $r$ values are assortative in nature.

## 3. Results & Discussions:

We have constructed the hydrophobic, hydrophilic and charged residues' networks for each of the eighty-five proteins. It has been observed that all the hydrophobic residues of the hydrophobic network for each of all the proteins do not form a single cluster. In general, they form one (in some cases more than one) giant cluster associated with small sub-clusters and isolated nodes. The same feature has also been observed for both of the hydrophilic and charged networks. Thus, all of the above three types of networks are sparse networks. On the other hand, when we consider all amino acids' network within a protein, the nodes (amino acids) do form a single cluster. We have also observed that in each of the eighty-five proteins, the total number of sub-clusters and isolated nodes of the hydrophobic network is smaller than that of the charged and hydrophilic networks. Only in one protein, the number of sub-clusters and isolated nodes of hydrophilic network is higher than that of hydrophobic network. However, charged networks of fifty-six proteins (out of eighty five proteins) show higher number of sub-clusters and isolated nodes than the respective hydrophilic networks. Thus we may say that hydrophilic and charged residues' networks within a protein are more sparse in nature than hydrophobic residues' networks.

To calculate and analyze the different network properties, we have selected those sub-clusters, which have at least thirty nodes. Thus we have finally obtained ninety-two hydrophobic, ninety-nine hydrophilic and sixty-nine charged sub-clusters with the criteria of having at least thirty nodes. We have further observed that the average number of nodes (amino acids) of hydrophobic sub-clusters is respectively nearly double and quadruple than that of hydrophilic and charged sub-clusters as is evident from Table 1.

It should be clearly mentioned that all the network parameters we have further calculated and analyzed are the result of our finally selected different sub-clusters or all amino acids' networks. In the rest of this paper, we would call these sub-clusters as networks.

### 3.1 Average degree of the networks:

For each of the four types of networks (BN, IN, CN and AN) we have calculated the average degree $<k>$. The values are listed in Table 1. We find that the average degree of hydrophobic networks ($<k^b>$), hydrophilic networks ($<k^i>$), charged networks ($<k^c>$) and all amino acids' networks ($<k^a>$) varies from 2.97 to 5.47, from 2.22 to 3.81, from 2.06 to 4.18 and from 6.75 to 10.09, respectively. The average of the $<k^b>$ values for all of the hydrophobic networks, $<k^b_{av}>$ was found to be 4.84 with a standard deviation 0.35. The average of the $<k^i>$ values for all of the hydrophilic networks, $<k^i_{av}>$ was found to be 2.97 with a standard deviation 0.29. For the charged networks, the average ($<k^c_{av}>$) was found to be 2.2.72 with a standard deviation 0.33.

It has been observed that the average of the $<k_a>$ of all of the all amino acids' networks, $<k^a_{av}>$ shows expected higher values than that of BN, IN and CN. Our results also clearly show that $<k^b_{av}> > <k^i_{av}> \approx <k^c_{av}>$. The Mann-Whitney U-test shows that these three populations are significantly different (level of significance is 0.001). To verify whether the observed trend is due to the network size or is purely the characteristic of the nature of the nodes of the network, we have compared the $<k>$ values of different networks with similar sizes (i.e. nearly same number of nodes). The result confirms the trend previously described. Hence our observation ($<k^b_{av}> > <k^i_{av}> \approx <k^c_{av}>$) is clearly an inherent nature of the network. We have also observed that within the same populations the value of average degree does not depend on the network size (i.e., on the number of amino acids of the protein).

### 3.2 Average strength of the networks:

Next we have studied the strength of the nodes within different types of weighted networks. The average strength of the hydrophobic networks $(<s^b>)$ varies from 17.28 to 35.21; whereas that of the hydrophilic $(<s^i>)$ and charged $(<s^c>)$ networks varies from 6.76 to 27.74 and from 14.71 to 50.63, respectively. On the other hand the average strength of AN

$(<s^a>)$ varies from 34.85 to 83.86. The average of $<s^a>$ for all of the AN networks was found to be 41.94 with a standard deviation 5.61. The average of the $<s^b>$ values for all of the hydrophobic networks, $<s^b_{av}>$ is nearly equal to that $(<s^c_{av}>)$ of the charged networks; whereas that of hydrophilic networks has smaller value than that of hydrophobic and charged networks.

### 3.3 Strength-degree relationships:

To understand the relationship between the strength of a node with its degree, $k$ we have further studied the average strength $<s^b>(k)$, $<s^i>(k)$ and $<s^c>(k)$ as a function of $k$. The result is shown in Fig 1. We have observed that the strength of a vertex changes with its degree, $k$. The average strength for all of the hydrophobic networks varies linearly with its degree, $k$. On the other hand, the average strength of charged and hydrophilic networks increases linearly with $k$ for smaller values of $k$, but sharply for higher values. It has been further noted that the slope of the best-fit line is different for different types of networks. The average strength of a node in charged networks increases more sharply than that of the hydrophobic and hydrophilic network as is evident from Fig 1.

### 3.4 Small World Property:

To examine whether the networks have the 'small world' property or not, we have calculated the average clustering coefficient $<C>$ and the characteristic path length $<L>$ for each of the networks and their respective values $(<C_r>$ and $<L_r>)$ for the random network having the same $N$ (number of nodes) and $<k>$. The average of the $<C>$ and $<L>$ values for all of the hydrophobic networks are given in Table 1. Those of hydrophilic and charged networks are also presented in Table 1. The ratios $[p=<C>/<C_r>]$ of average clustering coefficients of BN to that of classical random graph vary from 3.55 to 40.37. The ratios for IN and CN vary from 5.14 to 42.55 and from 3.69 to 24.32, respectively. On the other hand, it has been observed that the characteristic path length is of the same order as that of corresponding random graph as is evident from $q(=<L>/<L_r>)$ values listed in table 1. Although the ratios ($p$) for networks under study are not of the order of $10^2$ - $10^4$ as observed in the case scientific collaboration networks and networks of film actors, there are several other networks where $p$ may have smaller values (2,6,19,24,25). For example, the ratio ($p$) for metabolic network, protein-protein interaction network, food webs and network of *C. Elegans* has values 5.0, 4.4, 12.0 and 5.6, respectively. Even recent study on amino acid network within protein reported that the ratio ($p$) vary from 4.61 to 25.20 depending on the size of the network (18). Thus we may conclude that each of the three different types of networks (BN, IN and CN) has 'small world' property. We have also examined the all amino acids' network of the same proteins. We find that the ANs also have the 'small world' property as is evident from the $p$ and $q$ values listed in Table 1.

We have further studied the dependencies of $p$ and $q$ on $N$, number of nodes. The results are shown in Fig 2. We find that both the ratios $p$ and $q$ vary with $N$; but with different relationships. The ratio ($p$) of clustering coefficients varies linearly with $N$; whereas the ratio ($q$) of characteristic path lengths varies logarithmically with $N$. It should be mentioned that the $p$ values of ANs vary from 23.10 to 60.66. The higher $p$ values of ANs obtained in the present study than those as reported in (18) may be due to the larger size of networks.

### 3.5 Mixing behavior of the nodes:

We have also calculated Pearson correlation coefficients ($r$) for each of the networks. Almost all the hydrophobic networks (except one) have positive $r^b$ values and they vary from 0.02 to 0.43 with an average 0.30. While most of the hydrophobic networks have positive $r$ values, both of the hydrophilic and charged networks have both positive and negative $r$ values. The positive $r$ value of a network suggests that the mixing behavior of the nodes of that network is assortative type; whereas the negative $r$ value implies that the network is of disassortative type. The percentage of hydrophilic networks having negative $r$ values is significantly higher and lower than that of hydrophobic and charged networks respectively. When we consider the networks having non-negative $r$ values; the $r$ values of hydrophilic networks ($r^i$) vary from 0.00 to 0.52 and those of charged networks ($r^c$) vary from 0.00 to 0.51. The average of the $r^i$ values was found to be 0.20 with a standard deviation 0.10; whereas that of $r^c$ values was found to be 0.19 with a standard deviation 0.12. In case of all amino acids' network the $r^a$ values vary from 0.22 to 0.43. The average of the $r^a$ values was found to be 0.30 with a standard deviation 0.04.

The $r$ values of different networks suggest that the all amino acids' networks are of assortative type, the hydrophobic networks (except one) are also of assortative type. While maximum of the hydrophilic and charged networks are of assortative type, few others have the characteristics of disassortative mixing of the nodes as is evident from the $r$ values (data is not shown for negative $r$ values). Thus we may say that in almost all of the hydrophobic networks the hydrophobic residues (nodes) with high degree have tendencies to be attached with the hydrophobic residues having high $k$ values. Most of the hydrophilic and charged residues within their respective networks do follow the same behavior as followed by the hydrophobic residues. In a very few networks having negative $r$ values the mixing pattern of amino acid residues are different. Here the amino acids (nodes) having high $k$ values have a tendency to be attached with amino acids with smaller degree. A protein, in general, has hydrophobic, hydrophilic and charged residues. Thus an all amino acids' network is basically a composite network of these three types (hydrophobic, hydrophilic and charged) of networks. When we consider all amino acids' networks, we have obtained the $r$ values which represents a cumulative effect of either all positive $r$ values or a mixture of positive and negative $r$ values. Thus we find that the all amino acids' networks always have positive $r$ values.

### 3.6 Weighted and unweighted clustering coefficients of networks:

We have calculated the weighted and unweighted clustering coefficients of each of the hydrophobic, hydrophilic and charged residues' networks. The average clustering coefficients of hydrophobic, hydrophilic and charged residues' networks are assembled separately to make the ensemble of each type. The average of each of the ensembles has been calculated and is listed in Table 1.

In the present study, the unweighted clustering coefficients of hydrophobic networks vary from 0.41 to 0.55; whereas those of hydrophilic and charged networks vary from 0.38 to 0.63 and from 0.38 to 0.67, respectively. It is evident from Table 1 that $<C^b_{av}> < <C^i_{av}> < <C^c_{av}>$. We also find that the average weighted clustering coefficients of hydrophobic, hydrophilic and charged networks vary from 0.21 to 0.28, from 0.19 to 0.33 and from 0.19 to 0.34, respectively. We have also observed that $<C^{w,b}_{av}> < <C^{w,i}_{av}> < <C^{w,c}_{av}>$. The average weighted clustering coefficient is always nearly half than that of unweighted networks. In summary, the two major observations are (i) both the unweighted and weighted clustering coefficient values of hydrophilic networks are higher than those of hydrophobic, but are smaller than those of charged networks; and (ii) the average unweighted clustering coefficients are double than those of weighted clustering coefficients. The second observation indicates that the topological clustering is generated by edges with low weights. It further implies that the largest part of interactions (i.e. interactions between two amino acids) is occurring on edges (amino acids) not belonging to interconnected triplets. Therefore the clustering has only a minor effect in the organization of each of the three different

(hydrophobic, hydrophilic and charged) types of networks. On the other hand the unweighted clustering coefficient is a measure of local cohesiveness and the weighted clustering coefficient takes into account the strength of the local cohesiveness. Thus the first observation implies that hydrophilic networks have higher and lower local cohesiveness than hydrophobic and charged networks respectively.

### 3.7 Is there any hierarchical signature within the networks?

We have also studied the relationship of the clustering coefficients for both weighted and unweighted networks with their degree $k$. We find that for most of the hydrophobic networks having $k > 8$, both the unweighted $(<C^b>(k))$ and weighted $(<C^{b,w}>(k))$ clustering coeffcients change with their degree $k$. The results are plotted in Fig 3. It has been observed that the nodes with smaller $k$ values have higher clustering coefficients than the nodes with higher $k$ values. It is known that the hierarchical signature of a network lies in the scaling coefficient of $C(k) \sim k^{-\beta}$. The network is hierarchical if $\beta$ has a value of 1; whereas, for a non-hierarchical network the value of $\beta$ is 0 (6, 26). The low degree nodes in a hierarchical network belong generally to well interconnected communities (high clustering coefficients) with hubs connecting many nodes that are not directly connected (small clustering coefficient). Since in most of the hydrophobic networks, $C(k)$ significantly changes with $k$, we intend to study the possibility of hierarchy in the hydrophobic network. Here, both the $<C^b>(k)$ and $<C^{w,b}>(k)$ exhibit a power-law decay as a function of $k$ as is evident from Fig 3. It should be noted that we are aware of the problem in drawing conclusions about the power-law scaling and deriving exponents as well with such limited range of values. But this small range of $k$ values is actually a limitation of this real physical network. At the same time we have observed that both the $<C^b>(k)$ and $<C^{w,b}>(k)$ decrease significantly with $k$. So, it may be worthwhile to get an idea about the scaling coefficient values and hence, also about the nature of networks. However, the scaling coefficient ($\beta$) for the $<C^b>(k)$ varies from 0.005 to 0.750 with an average of 0.254; whereas the corresponding coefficient ($\beta^w$) for $<C^{w,b}>(k)$ varies from 0.025 to 0.755 with an average of 0.231. We observe a power law decay for both $<C^b>(k)$ and $<C^{w,b}>(k)$, but the average values ($\beta$ and $\beta^w$) of the scaling coefficients lie neither very close to 0 or 1 but take intermediate values. The values of the scaling coefficients imply that the networks have a tendency to hierarchical nature.

But the unweighted and weighted clustering coefficients of both the hydrophilic and charged residues do not show any clear functional relationship with their degree $k$ as is evident from Fig 3. We have already mentioned that the small range of $k$ values impose a problem in drawing conclusions about the power-law scaling and deriving its exponents. In spite of the limitations, we may say that the hydrophobic networks bear the signature of hierarchy; whereas the hydrophilic and charged networks do not have any hierarchical signature. We have further observed that all amino acid residues' network exhibits a signature of hierarchy as is evident from Fig 3 and from the values of scaling coefficients listed in Table 1. The same observation has also been mentioned in (18). Thus we may say that the hierarchical signature of all amino acids residues' network is mainly originated from the hierarchical behavior of hydrophobic residues' network.

### 3.8 Degree and Strength distribution:

We have also studied the probability degree and strength distributions of AN, BN, IN and CN. We have observed that the probability degree distribution of network connectivities of all four types of networks (AN, BN, IN and CN) has a peak followed by a decay whose exact nature is difficult to determine because of the small number of $k$ values (data not shown). On the other hand the probability strength distributions exhibit a large number of fluctuations (data not shown). It makes difficult to find the exact nature of the distributions.

**4. Conclusion:**

In summary, all of three types of networks (hydrophobic, hydrophilic and charged) as well as all amino acids networks have the 'small world' property. While hydrophobic, hydrophilic and charged residues' networks are sparse in nature, all amino acids' networks do not have any sub-clusters or isolated nodes. The total number of sub-clusters and isolated nodes in hydrophobic networks of each of the proteins we have studied is significantly smaller than that of hydrophilic and charged networks. The average degree of hydrophilic and charged networks has significantly smaller value than that of hydrophobic networks. On the other hand, the average strength of the hydrophobic and charged networks has higher value than that of hydrophilic networks. We have also observed that the average strength of the charged networks is nearly equal to that of hydrophobic networks. While the average strength of the nodes (residues) for each of the three types of networks (BN, IN and CN) varies with its degree, $k$; the average strength of a node in charged networks increases more sharply than that of hydrophobic and hydrophilic networks. We have further observed that all amino acids' networks and hydrophobic networks are of assortative type. While maximum of the hydrophilic and charged networks are of assortative type, few others have the characteristics of disassortative mixing of the nodes. We have also observed that all amino acids' networks and hydrophobic networks bear the signature of hierarchy; whereas the hydrophilic and charged networks do not have any hierarchical signature.

**Acknowledgments:**

The authors acknowledge the computational support provided by Distributed Information Center of Calcutta University.

TABLE 1: Different network properties - average number of nodes (<$N_r$>), average degree (<$k$>), average strength (<$s$>), average characteristic path length (<$L$>), average clustering coefficients of unweighted (<$C$>) and weighted (<$C^w$>) networks, Pearson correlation coefficient *(<$r$>)*, the average ratios *(<$p$> and <$q$>)*, average scaling coefficients of unweighted *(<$\beta$>)* and weighted (<$\beta^w$>) networks – of hydrophobic (BN), hydrophilic (IN), charged (CN) and all amino acids (AN) networks.

| Network Type | <$N_r$> | <$k$> | <$s$> | <$L$> | <$C$> | <$C^w$> | <$r$>* | <$p$> | <$q$> | <$\beta$>† | <$\beta^w$>† |
|---|---|---|---|---|---|---|---|---|---|---|---|
| BN | 221.22 ±73.29 | 4.84 ±0.35 | 23.72 ±2.74 | 7.45 ±1.59 | 0.46 ±0.02 | 0.23 ±0.01 | 0.30 ±0.07 | 20.76 ±6.71 | 2.18 ±0.34 | 0.254 ±0.125 | 0.231 ±0.124 |
| IN | 92.78 ±56.08 | 2.97 ±0.29 | 14.59 ±2.95 | 7.96 ±2.38 | 0.49 ±0.05 | 0.25 ±0.02 | 0.20 ±0.10 | 14.98 ±8.22 | 1.94 ±0.40 | | |
| CN | 45.97 ±18.60 | 2.72 ±0.33 | 22.46 ±5.59 | 6.73 ±1.74 | 0.52 ±0.06 | 0.27 ±0.03 | 0.19 ±0.12 | 8.67 ±3.12 | 1.74 ±0.32 | | |
| AN | 612.15 ±134.82 | 7.58 ±0.38 | 41.94 ±5.61 | 6.61 ±0.88 | 0.37 ±0.07 | 0.19 ±0.01 | 0.30 ±0.04 | 29.71 ±6.46 | 2.09 ±0.22 | 0.208 ±0.110 | 0.166 ±0.106 |

*Data shown only for +ve <$r$>

†Since there is no clear functional relationship in case of IN and CN the values of scaling coefficients are not listed.

Figure Legends

FIG. 1: Average strength $\langle s \rangle(k)$ as a function of degree $k$ of hydrophobic (BN), hydrophilic (IN), charged (CN) and all amino acids' (AN) networks.

FIG. 2: The ratios $p\ (=\langle C \rangle/\langle Cr \rangle)$ and $q\ (=\langle L \rangle/\langle Lr \rangle)$ as a function of network size $N$. The ratio $p$ varies linearly with $N$; whereas the ratio $q$ varies logarithmically with $N$. The best-fit curves are shown by lines for the two ratios. (a) All amino acids' networks (AN), (b) Hydrophobic networks (BN), (c) Hydrophilic networks (IN) and (d) Charged networks (CN).

FIG. 3: Topological clustering coefficient $C(k)$ and weighted clustering coefficient $C^w(k)$ as a function of degree $k$ for different types of networks [(a) All amino acids' networks(AN), (b) Hydrophobic networks(BN), (c) Hydrophilic networks(IN) and (d) Charged networks(CN)] for a representative protein (PDB Id:8ACN). The best-fit curves are shown by lines.

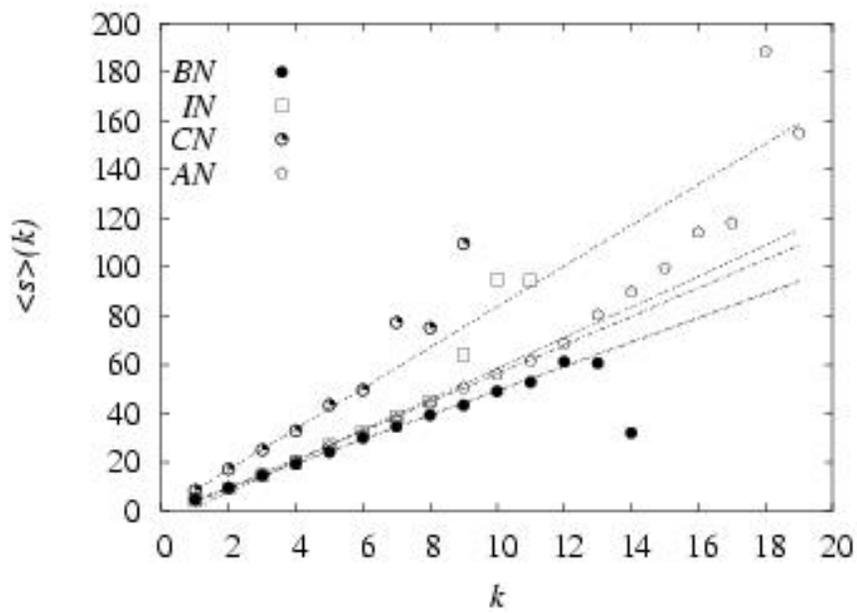

FIG 1

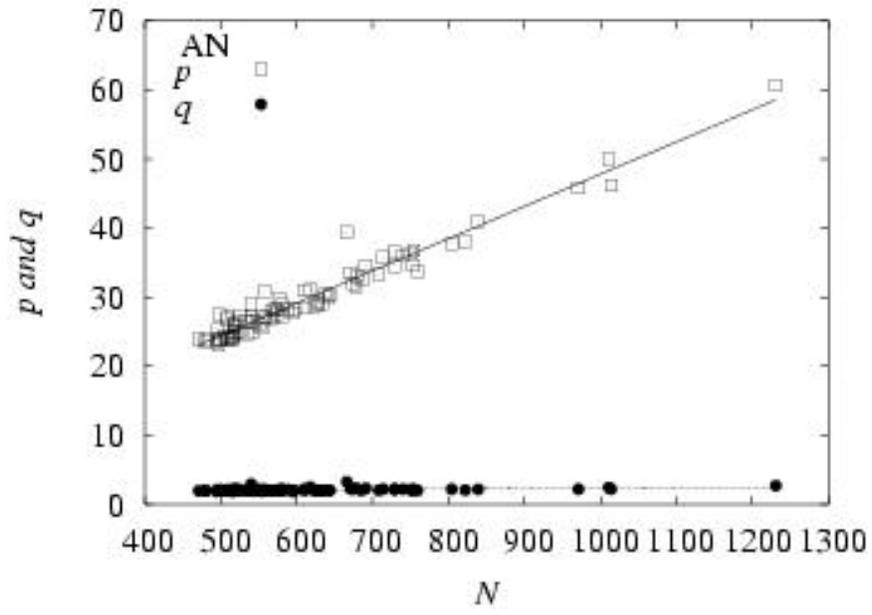

FIG. 2(a)

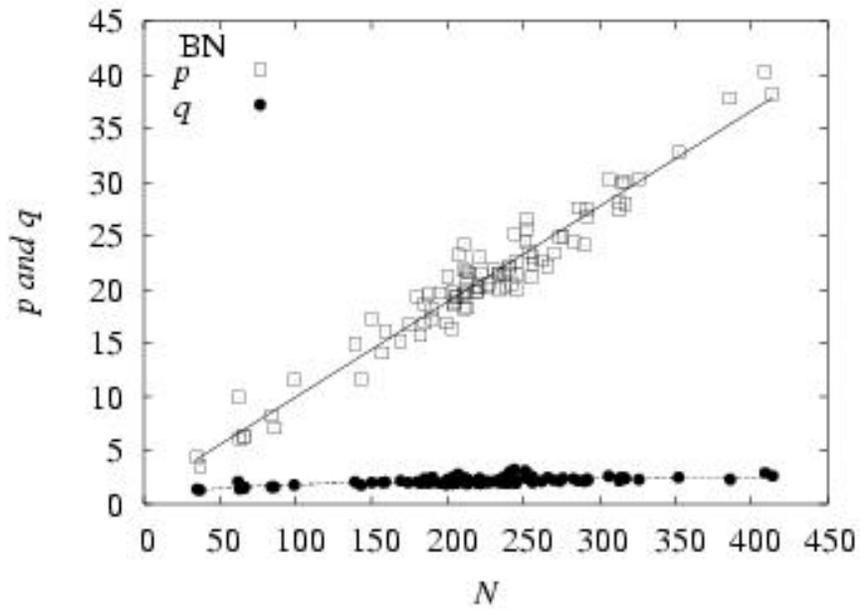

FIG. 2(b)

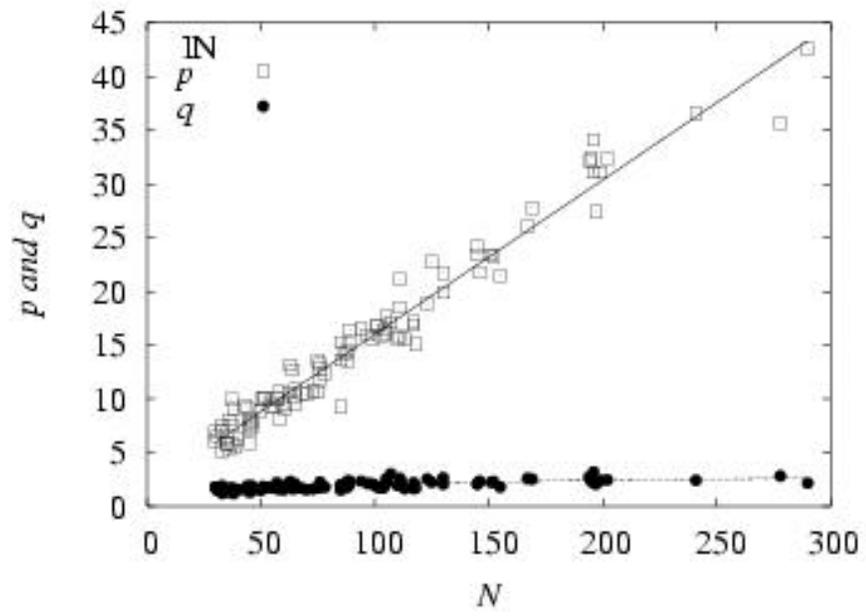

FIG. 2(c)

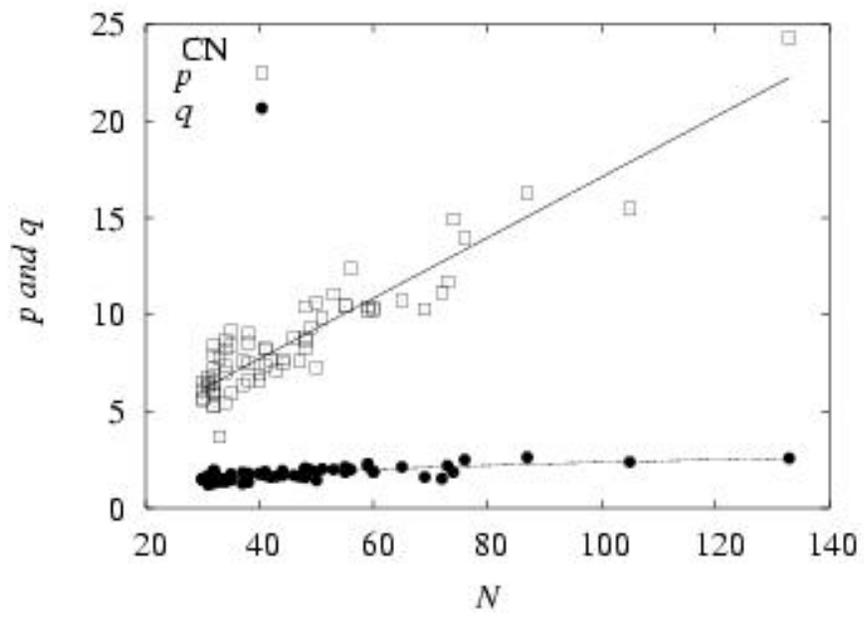

FIG. 2(d)

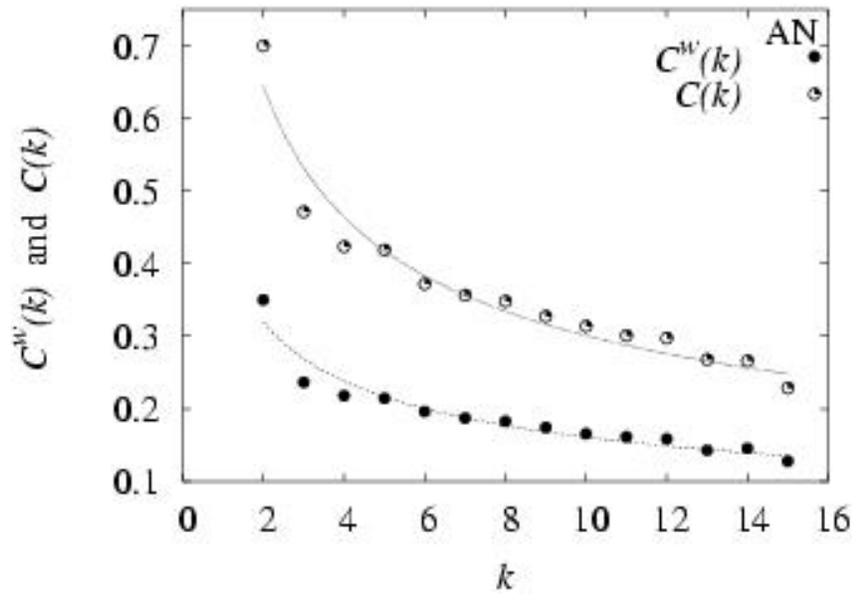

FIG. 3(a)

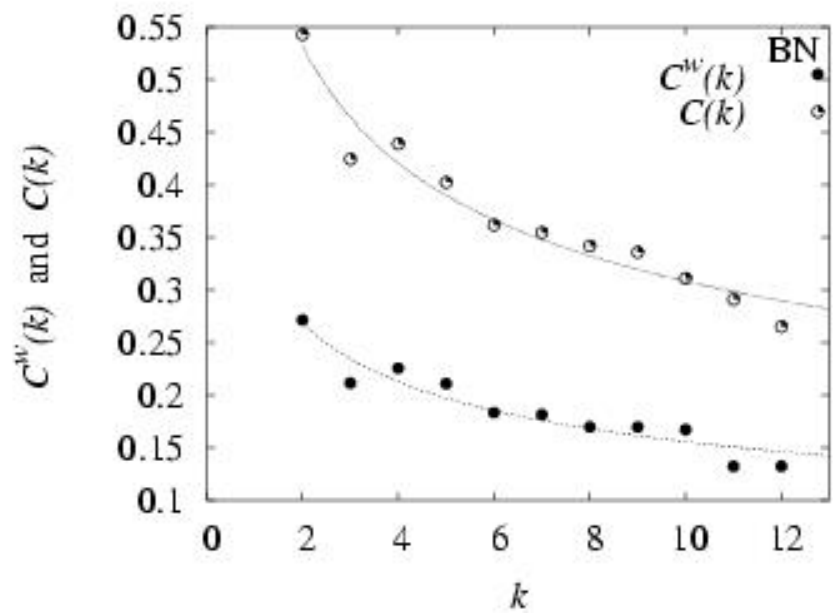

FIG. 3(b)

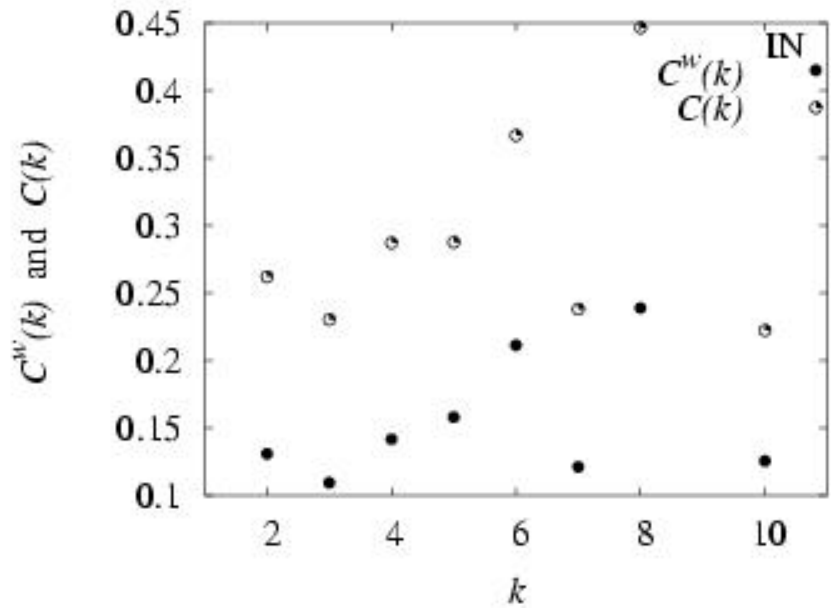

FIG. 3(c)

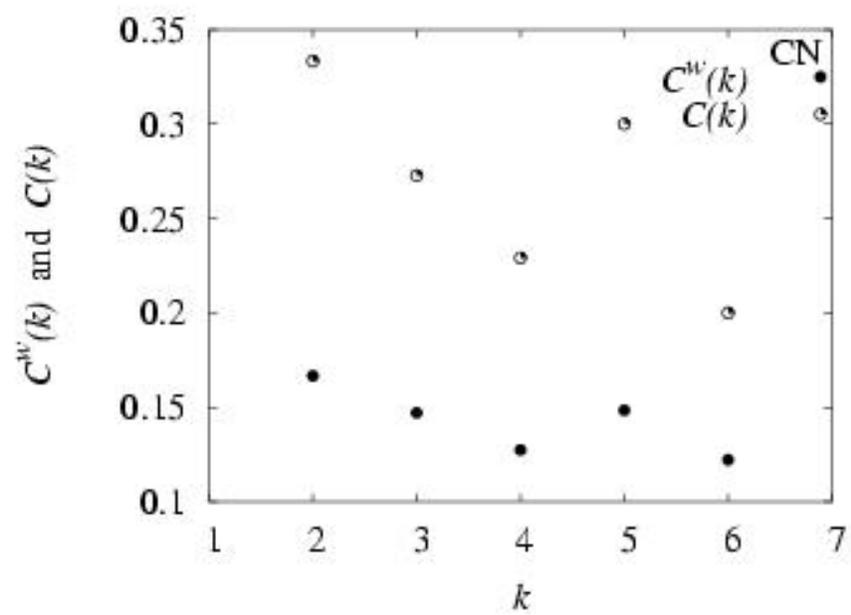

FIG. 3(d)